\documentclass[10pt,twocolumn, twoside]{IEEEtran}
\usepackage{tabularx}
\usepackage{graphicx}
\usepackage{subfigure}
\usepackage{color}
\usepackage[utf8]{inputenc}
\usepackage{amsmath,amssymb,epsfig,theorem,url,cite,bm,soul}
\usepackage{multirow}
\usepackage{rotating}
\usepackage{wrapfig }
\usepackage{times}
\usepackage{fancyhdr,graphicx,amsmath,amssymb}
\usepackage[ruled,vlined]{algorithm2e}
\include{pythonlisting}

\usepackage{soul}
\usepackage{color}
\usepackage[usenames,dvipsnames]{xcolor}

\author{\IEEEauthorblockN{Enes Altinisik, 
H\"usrev~Taha~Sencar 
}

}

\begin{document}

\title{Automatic Generation of H.264 Parameter Sets to Recover Video File Fragments}
\maketitle

\begin{abstract}
We address the problem of decoding video file fragments when the necessary encoding parameters are missing.
With this objective, we propose a method that automatically generates H.264 video headers containing these parameters and extracts coded pictures in the partially available compressed video data.
To accomplish this, we examined a very large corpus of videos to learn patterns of encoding settings commonly used by encoders
and created a parameter dictionary.
Further, to facilitate a more efficient search our method identifies characteristics of a coded bitstream to discriminate the entropy coding mode. 
It also utilizes the application logs created by the decoder to identify correct parameter values.
Evaluation of the effectiveness of the proposed method on more than 55K videos with diverse provenance shows that it can generate valid headers on average in 11.3 decoding trials per video.
This result represents an improvement by more than a factor of 10 over the conventional approach of video header stitching to recover video file fragments. 

\end{abstract}

\section{Introduction}
\label{sec:Intro}

Recovery of the content of a file when it is only partially available is a challenging task especially for multimedia data.
Crucially, a file is the basic unit of information organization in all computing systems.
The syntax of the data in a file is specified based on the requirements of the application using it, and the data is mostly encoded and compressed
to ensure efficient  storage and transfer.
For these reasons, access to file data is governed by the use of a decoder which validates the file's syntax, interprets it, and reconstructs the raw data. 
As the complexity of encoding methods, such as those used for compressing images and videos, increases, a large number of encoding parameters have to be set to guide the decoding process.
These parameters are encapsulated within what is widely referred to as the file-header. 
Without this header decoding cannot be performed even if the rest of the file data is intact.

File data is always stored and transferred in fixed-size blocks, such as disk sectors, memory pages, SSD blocks, and network packets.
Since a file-header comprises a very small part of the overall file data, there are several cases in which it may be missing.
In digital forensics, extraction of digital evidence from systems through a process known as file carving is a standard procedure.
In this context, partial files are frequently encountered when carving partly deleted files or when file system metadata is not available due to corruption or hardware failures.
Similarly, in the network monitoring domain deep packet inspection (DPI) is a technique used for analyzing network and application traffic at the packet level.
In very high-speed data links, errors in the form of dropped packets are inevitable, making it difficult to extend DPI capabilities to multimedia data in general.
Due to the prevalence of videos, in both problem domains, accessing partial video file data, when the header is not available, is an important requirement. 

In fact, the video file carving problem has long been the focus of research.
Since video files are typically very large, they are very likely to be partitioned into a large number of fragments on the storage \cite{garfinkel2007carving}. 
Motivated by this fact, the main research effort has focused on the video defragmentation problem where the goal is to reorder and combine fragments of a file to reconstruct the original videos. 
The main objective of this work is to essentially reassemble the fragmented file as correctly as possible while minimizing the computational cost. 
To this objective, several approaches utilizing different adjacency metrics \cite{poisel2011advanced, alghafli2019}, 
clustering methods to discriminate fragments associated with different videos \cite{twoStage}, and the use of other file metadata (such as camera ID, location, timestamp, {\em etc.}) in frame headers \cite{lewis2012reconstructing,park2014data} have been proposed.
More critically, one thing these file carving approaches have in common is that they assume the availability of file headers.

Since video data has always been stored in some compressed form, a more important and essential capability is the recovery of video content when the file headers needed for decoding are missing.
In fact, in many real-world settings, making an arbitrary video fragment playable sufficiently addresses the need. 
Further, such a capability significantly reduces the complexity of reassembling fragmented videos as each fragment can be evaluated individually based on its content, without the need for repetitive fragment stitching and decoding to validate the adjacency of fragments.
This, however, is a more challenging problem as it requires blindly determining decoding parameters associated with a given file fragment. 

To access a video file, the decoder first needs to initialize itself using parameters available in the header. 
Since video coding involves a large number of parameters, performing a brute force search in the whole parameter space is not feasible. 
So far, only a very few studies have aimed at addressing this problem.
The earliest work in this direction is a carving method, Defraser, an application introduced by the Netherlands Forensic Institute (NFI) \cite{NFI}, 
that essentially uses a library of headers extracted from previously acquired, intact H.264 coded videos and tries to decode a fragment using one of these reference headers\footnote{Defraser is a commercially available video carving tool. It enables users to populate a header database.}.
In \cite{na2013frame}, Na {\em et al.} analyzed MP4 file format \cite{pereira2002mpeg} (a container file for H.264 coded video) and the H.264 Annex-B \cite{tsbmail}, which standardizes the bitstream format for H.264 encoding, to identify patterns that mark the beginning of coded video frame data. 
They utilized this information to alleviate the defragmentation problem by searching for frame data and appending a known or reference header, similar to Defraser, to decode the frame data.  
In \cite{yannikos2013automating}, considering MPEG-1 coded videos, Yannikos {\em et al.} introduced another reference header-based approach that also estimates the picture width to alleviate the search complexity.  

An important limitation of using reference headers is that it potentially results in a very large search space as it cannot discriminate 
between parameters in terms of how critical they are for decoding.
Further, it is hard to span all the possible headers used in practice. 
Smartphones and tablets have become the standard camera today, and they support several video recording settings that may be selected through system-provided interface \cite{ei2019}. 
That is, the headers of videos captured by one camera app will expose only certain encoding settings and do not reveal all possible options that may potentially be selected by other camera apps that utilize the built-in camera hardware in a device.
More importantly, these parameters need not be fixed throughout the video as encoders dynamically determine them during encoding to achieve a target bitrate or quality.
Therefore, a more generalizable approach is to identify individual parameters needed to decode a file without a failure, as it can dramatically reduce the header search space.
In fact, examining the parameters that comprise H.264 video headers, Sheng {\em et al.} \cite{reconstructHeader} identified 13 parameters and noted that only three of them are critical for decoding.

In this regard, our approach is most similar to earlier work focusing on recovery of JPEG file fragments that took a data-driven approach to 
learn encoding settings of more than 3.2K camera models and several photo editing tools in order to render the partial image from a block of JPEG coded data \cite{uzun2015carving, uzun2019jpg}.
However, there are important differences in the challenges that need to be addressed between recovering image and video file fragments.
Most notably, in the former the challenge arises from working at the subframe level, requiring the ability to decode an entropy coded sequence from an arbitrary point. By contrast, when recovering video file fragments this is of little concern, as typically there  are  many  intact  entropy coded frames available for decoding. 
The main challenge here is due to the more advanced nature of video coding which results in the use of a very large number of parameters in comparison to  image  coding.  
Hence,  the  crux  of  the  problem lies in efficient search of an extremely large parameter space.

Our method of search for decoding parameters utilizes both domain knowledge and observations of the prevalence of encoding parameters used in practice by several cameras.  
Our analyses show that parameters can be categorized into three groups based on how they must be treated.
To reduce the parameter search space, we further rely on learning approaches 
that utilize errors encountered during decoding in response to predefined settings
and bitstream characteristics.
Our approach is designed and validated on a large dataset, including more than hundred thousand videos, the largest used in any study of this nature.
Results show that our method of header generation is able to determine the encoding parameters on average in 11.3 decoding trials.
Our findings on video encoding settings also add a new dimension to the research effort in file metadata-based video source identification \cite{gloe2014forensic,iuliani2018video,lopez2020digital,huaman2020authentication} and tampering detection \cite{song2016integrity,iuliani2018video,guera2019we,huaman2020authentication}.

In the next section, we describe the defining characteristics of video files together with statistics obtained from our collection of videos. 
This is followed by an overview of how an H.264 coded video sequence is generated along with a description of parameters driving the coding process in Section \ref{sec:H264}. 
Section \ref{sec:Prevalence} categorizes parameters into three groups based on encoding settings observed in actual videos and the results of analyses performed by actively changing the coding settings of these videos.
Our header-generation method is described in Section \ref{sec:generation}, and an evaluation of its performance on a large set of videos is presented in Section \ref{sec:Evaluation}.
\textcolor{black}{Finally, Section \ref{sec:Discussion} provides a discussion of our results, and Section \ref{sec:Conclusion} concludes and points to possible directions for future work.}

\section{Video Files in Practice} 
\label{sec:Practice}
The creation of a video file is defined by two processes. 
The first is the video encoding which reduces the temporal and spatial correlations in the original sequence of frames by predicting frame regions using other visually similar regions.
Starting with the introduction of H.261 in 1988, the first practical video coding standard based on the use of discrete cosine transform (DCT), several international video coding standards have been developed based on its design and standardization procedure.  
These include MPEG-1 \cite{mpeg1} in 1991, MPEG2/H262 in 1994, MPEG4/H263 \cite{H263} in 1999, and MPEG4 AVC/H264 \cite{H264} in 2003.
More recently, in 2013, H.265 coding standard \cite{H265} was introduced in response to increasing resolution and quality of videos with parallel processing in mind. 
In addition to these, many platform-driven coding formats inspired by the designs of H.264 and H.265 standards, such as VP8 and VP9 developed by Google, Apple ProRes developed by Apple, VC developed by Microsoft as well as the AV1 format introduced by the Alliance for Open Media are also available for use.  
In this diversity of codecs, H.264 is reported to be the most commonly used coding standard comprising 91\% of all videos \cite{bitmovin}.
This can be largely attributed to the fact that parts of the technology underlying H.265 coding are patented and subject to licensing fees.  

The other defining characteristic of a video file concerns the encapsulation of the encoded data with other essential data, such as coded audio data, encoding parameters, subtitles, and other metadata in a container file.
Several types of container formats are available for both streaming and storage of videos, such as MP4 format based on MPEG-4 Part 14 standard, QuickTime (MOV) format developed by Apple, and AVI and WMV formats developed by Microsoft. 
Essentially, container formats are optimized for different use cases in the way they organize the media data and the range of audio and video codecs they support.  
In this regard, MP4 is one of the most widely used container formats due to its versatility and is also used by streaming services such as YouTube and Vimeo.

Since the container is a wrapper for encoded data, recovering video content from a partial video file ultimately depends on the ability to identify and decode the
coded video frames. 
To examine video encoding characteristics, we built a comprehensive dataset of videos derived from publicly available sources.
For this, we crawled the decentralized content sharing and publishing platform \texttt{lbry.tv} \cite{lbry}. 
Unlike many other well-known video sharing platforms, the \texttt{lbry.tv} platform does not re-encode the videos uploaded by its users.
That is, the published videos are not recompressed\footnote{https://lbry.com/faq/video-publishing-guide}. 
This potentially provides us with a collection of videos recorded using many different cameras and processed by several video editing tools.  
The crawling process took place in two separate periods to obtain videos needed for the design and test phases of our study.   
Overall, we downloaded a total of 102,846 publicly accessible videos shared by 16,874 individual users.

\begin{figure}[htbp]
\centering
    \centering
    \includegraphics[width=0.75\columnwidth]{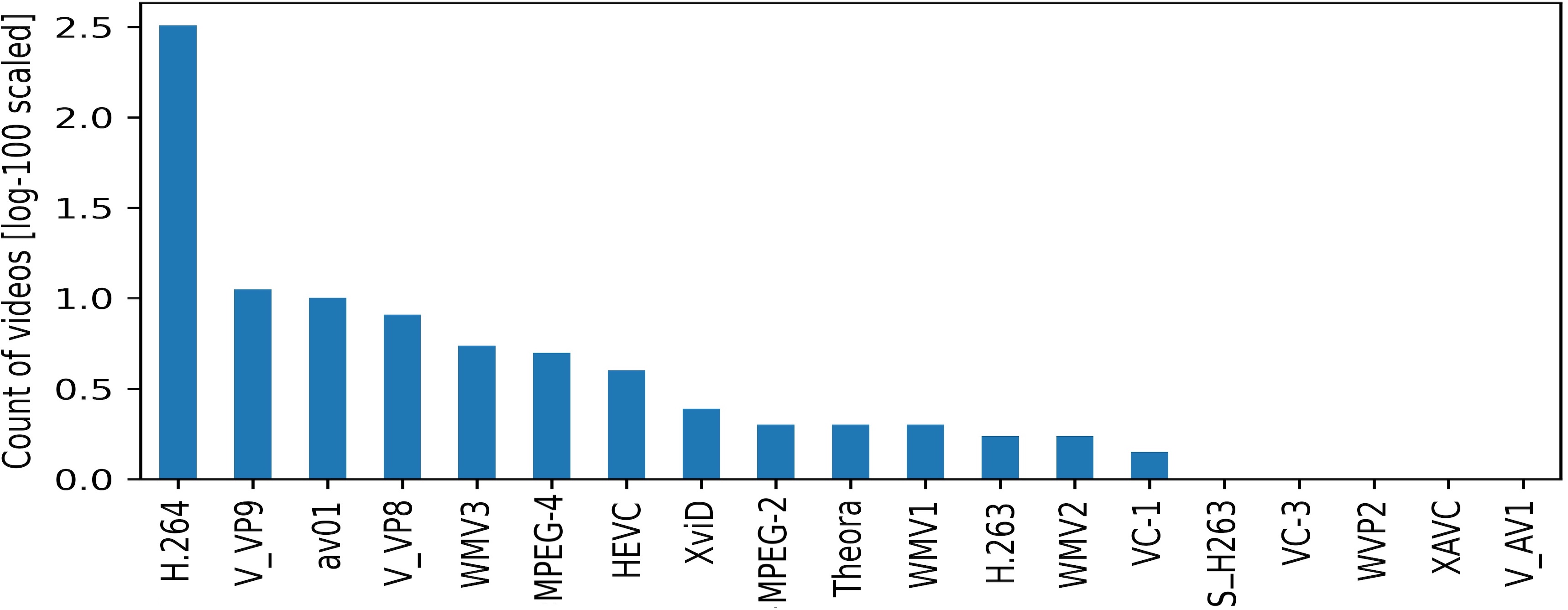}
  \caption{Distribution of encountered video encoding formats in our dataset.}
  \label{fig:dist}
\end{figure}

This dataset is then further enhanced by two public video datasets created for facilitating the study of the source camera identification problem.
The first is SOCRatES \cite{socrates} which includes 1,000 videos captured by 104 different smartphone cameras of 15 different makes and 60 different models. 
The other dataset, VISION \cite{vision}, includes 648 videos recorded using 35 smartphones of 11 major brands.
Overall, this resulted in a dataset of 104,524 videos in MP4 container file format.
The distribution of encountered encoding formats in our dataset is displayed in Fig. \ref{fig:dist}.
As can be seen, 99.6\% of all videos, corresponding to 104,139 videos, are encoded using H.264 format.
Due to this finding, we decided to tailor our recovery method to only take into account H.264 encoding and used the corresponding 104,139 videos for building and evaluating our header-generation approach.

\section{Organization of H.264 Coded Data Stream}
\label{sec:H264}
The H.264 standard organizes video coding functions into two conceptual layers. 
The Video Coding Layer (VCL) governs the encoding process and includes all functions related to the compression of video frames. 
The second layer, referred to as Network Access Layer (NAL), involves encapsulation of the encoded data for efficient storage and transfer. 
Essentially, an H.264 bitstream contains a sequence of NAL units which serve as the building blocks of the H.264 stream.
Therefore, from a data recovery point of view, the goal is to identify, extract, and interpret data in these units when some are missing.               

Every NAL unit includes a one-byte header where the trailing 5-bits specify the type of unit followed by a sequence of bytes called the raw byte sequence payload (RBSP).
The NAL units are byte aligned and are separated from each other by a prefix, referred to as start code.  
The payload of NAL units can either include encoded video data provided by the VCL or some additional information needed by the decoder.
Most critically, the latter type of payload includes  the Sequence Parameter Set (SPS) and the Picture Parameter Set (PPS). 
The SPS NAL unit contains parameters that are common among a series of consecutive coded video pictures.
The PPS NAL unit further complements the SPS by specifying parameters that apply to decoding of one or more pictures. 
Unlike the other NAL units that contain non-VCL payload, without these two units, which will be referred to as SPS and PPS headers, video frames cannot be decoded. 

The SPS and PPS information is followed by the coded video pictures that can be contained within several types of NAL units. 
Among these, the Instantaneous Decoder Refresh (IDR) picture is the most important as it marks the earliest frame that can be referenced during decoding 
and corresponds to an updated SPS and PPS.  
The IDR and non-IDR pictures can be divided into multiple slices during coding and each slice is contained within a separate NAL unit.
Overall, all IDR NAL units and the following non-IDR units can be decoded successfully as long as the corresponding SPS and PPS units are available.

\subsection{Placement of SPS and PPS NAL Units}

Ideally, it is possible to include a single SPS and PPS at the start of a video data stream.
However, several use cases dictate their repetition. 
The byte stream format described in the standard (Annex B format) inserts SPS and PPS headers before every IDR picture and PPS header before non-IDR pictures. 
In the case of video streaming, this is preferable as it allows a decoder to start decoding midstream. 
In addition, the encoder may vary parameters in different parts of the stream to achieve a target bitrate or quality.

Many video files are, however, intended for download and storage. 
Therefore a more favorable approach is to store one copy of each unique SPS and PPS unit in some part of the file, especially if they remain fixed for the whole video stream.
Video file containers tend to use this approach because of its efficiency.
An important downside of this approach is that when SPS and PPS headers cannot be located, decoding cannot be performed. 
The uncoupling of parameters from the coded frame data is the main motivation for our approach
as it requires blindly determining the parameters used during encoding.

\subsection{What Is in SPS?}
The SPS contains parameters that apply to a sequence of pictures, including one IDR picture and many non-IDR pictures.
The details of the SPS are provided in Table \ref{SPS-PPS-Headers} of the Appendix.
In essence, these parameter values declare the needed capabilities for decoding the video stream and initialize parameters that vary at the sequence level.

At a higher level, SPS includes four groups of parameters. 
The first group specifies the required capabilities for the decoder. 
These essentially allow the decoder to decide whether it can support the video encoding settings such as resolution, bitrate, and frame rate.
The second group of parameters defines the sequence properties, such as bit lengths of several variables and the number of frames that must be stored in the reference buffer. 
The decoder uses these parameters when parsing the bitstream and for memory management.
Resolution-related parameters comprise another group.
These parameters set the width and height values at a 16-pixel granularity in accordance with the size of a macroblock. 
If the width or height is not a multiple of 16, the remaining portions are defined in a number of frame-cropping parameters.
The last group does not directly affect encoding. 
It involves parameters like the ID assigned to each SPS which can be changed freely as long as it is kept consistent within the sequence. 
Similarly, the optional video usability information (VUI) parameters are only involved in the post-decoding stage when generating the video, after individual pictures are reconstructed.

\subsection{What Is in PPS?}

For every picture in a sequence, there may be a separate PPS specifying the encoding parameters of the picture, as listed in Table \ref{SPS-PPS-Headers} of the Appendix. 
The most important of these is the entropy coding mode, designating which of the two methods, namely the Context-Adaptive Binary Arithmetic Coding (CABAC) or Context-Adaptive Variable-Length Coding (CAVLC), is used for losslessly compressing coded picture data. 
In addition, several parameters define the structure of the slice groups for a picture and the mode of motion prediction. 
Another subgroup of parameters set the default quantization values related to macroblock data in picture slices. 
Finally, one parameter determines if the deblocking filter is applied at its default setting or using a custom setting with involved parameters specified in another NAL unit.

\section{Prevalence of Coding Parameters in Practice}
\label{sec:Prevalence}

The possible range of values for each parameter contained in SPS and PPS are given in Table~\ref{SPS-PPS-Headers}. 
Accordingly, the space of valid SPS and PPS units is extremely large. 
This crucially prohibits an exhaustive search over all possible parameter values. 
Nevertheless, the choice of SPS and PPS parameter values strongly depends on encoder implementations.
Since many cameras and editing software may assume similar settings, parameters may also not vary significantly among videos.
As an example, in the case of JPEG file format, which is the de-facto coding format used by digital cameras to save photos, an examination of more than seven million photos captured by 3,269 camera models revealed that 77\% of photos were using the sample Huffman code table sets given in the standard document instead of customizing them \cite{uzun2019jpg}.
Furthermore, for a given encoding setting, it is possible that some parameters may accept multiple values without impairing decoding.
Therefore, in practice, the actual space for parameter values used in encoding of videos can be expected to be smaller than what a brute-force search may require.

To estimate the effective search space for parameters and determine the most prevalent parameter values, we examined SPS and PPS NAL units extracted from videos in our dataset.
For this purpose, we divided our dataset into design and test partitions.
The design set includes a total of 48,118 obtained by combining VISION and SoCRATES datasets with videos acquired from 11,034 \texttt{lbry.tv} user accounts. 
The remaining 56,021 videos obtained from 5,840 non-overlapping and 3,979 overlapping \texttt{lbry.tv} user accounts are reserved for an evaluation study.
Examination of the videos in the design set yielded 5,115 unique SPS and PPS NAL unit combinations (7,383 when VUI parameters are also considered), comprising 4,180 different SPS units and 471 different PPS units.

To further investigate the criticality of each parameter for decoding, 
we performed an analysis by varying these parameter values and observed their impact on decoding.
To this objective, considering each SPS and PPS NAL unit combination encountered in the design set, we randomly selected one video encoded using those
parameter sets.
Then, each parameter in the combined parameter set is individually modified by substituting it with other parameter values observed in the remaining SPS and PPS pairs.
After each modification, decoding is performed using the modified NAL units, which differ from their original settings by only a single parameter's value\footnote{We used the H264 BitStream toolbox~\cite{h264bitstream} to edit SPS and PPS NAL units and FFMPEG video editing tool to perform decoding.}.
We determined that if a modified SPS and PPS cannot decode the first picture in a video sequence, the decoder will produce several error messages and will continue to try decoding subsequent pictures until no data is available, eventually creating an empty frame or an incorrect picture.

To identify parameter values that yield successful decoding, we compared the resulting frames with frames obtained from original videos.
We must note here that successful decoding requires correctly determining the picture width.
Otherwise, decoded picture blocks are misplaced. 
Since earlier decoded blocks serve as references for predicting subsequent blocks, block misplacement eventually causes a failure when the decoder cannot locate a needed block.  
Unlike the width, an incorrect picture height causes either picture cropping or picture elongation through content repetition.
Therefore, the correlation between an original picture and the reconstructed version can be used as a statistic to assess correctness of parameters.  
We observed that for correct decoding measured correlation values are always above the value of 0.8, and used it as a threshold to automate our analysis.

Based on these decoding steps, we determine how mismatching parameter settings are compatible with each other.
Our analysis on the design set essentially showed that parameters can be divided into three groups as identified in the second column of Table~\ref{SPS-PPS-Headers}.

\subsection{Core Parameters}
The first group includes parameters that are observed to vary significantly across videos and cause a failure when set incorrectly.
We identified 10 parameters in SPS and PPS in this category, as listed in the upper part of the Table~\ref{tab:parameters}.
These parameters can be further divided into two in terms of how they affect decoding (last column of the table). 
The first sub-group involves parameters, such as the type of entropy coding, bit lengths of several variables, and the presence of deblocking filter configuration, that cause a bitstream parsing error.  
When these parameters do not match their values used for encoding, the decoder cannot interpret the coded data in a meaningful way and a failure is imminent.
The second sub-group leads to an impossible-to-satisfy condition mainly due to quantization parameters and image dimensions being inconsistent with the decoded data.  
When they are incorrect, the decoder is very likely to fail at reconstructing video pictures.

With the latter sub-group, the ambiguity arises due to behavior induced by the three parameters.
For example, the quantization parameter which determines the level of quantization applied to picture macroblocks may take 50 values.
To improve coding efficiency, the value needed for reconstructing each coded picture macroblock is not stored as an absolute value.
Instead, the SPS sets a base value and the quantization parameter of each block is stored in fewer bits as either an increment or decrement with respect to this base value.
Hence when an incorrect base value is assumed, this may result with a non-valid (out-of-bound) quantization parameter value.
A similar phenomenon occurs for the cropping flag and right cropping offset parameters.
Overall, incorrectly setting these parameter values causes a decoding failure most of the time but not necessarily always.

These 10 parameters can be selected independently from each other and yield around 3.5 billion combinations, as indicated by the fourth and fifth columns of Table \ref{tab:parameters}.
In the design set, however, we encountered around 2,663 different combinations (10-tuples) of those core parameters, which verifies our intuition that certain encoding settings are more prevalent. 
Further, some parameters are determined to be interdependent. 
For example, in all videos the bit depths of luma and chroma samples are seen to be set to the same value, and the H.264 decoder implementation of FFMPEG video processing tool did not allow setting them otherwise.
Finally, we must also note that in \cite{reconstructHeader} only three parameters are identified to be crucial for decoding. 
Our observations on the design set in contrast show that the assumption of this work does not hold in practice. 
Similarly, the intuition behind the picture width estimation approach proposed by \cite{yannikos2013automating} does not apply to H.264 encoding due to deployment of inter- and intra-block prediction.

\vspace{-0.1cm}

\begin{table}[!ht]
	\centering
	\caption{Parameters That Must Be Identified for Decoding}
	\label{tab:parameters}
	\resizebox{\linewidth}{!}{
    	\begin{tabular}{|c|c|c|c|c|c|}
    	\hline
    	& \multirow{2}{*}{\shortstack{NAL\\Unit}} & \multirow{2}{*}{\shortstack{Parameter Name}} & \multicolumn{2}{|c|}{\shortstack{\# of Values}} & \multirow{2}{*}{\shortstack{Induced\\Error}} \\
    	 & & & Seen & Possible & \\\hline
    	 \parbox[t]{2mm}{\multirow{10}{*}{\rotatebox[origin=c]{90}{Variant}}} 
    	 & \parbox[t]{2mm}{\multirow{6}{*}{\rotatebox[origin=c]{90}{SPS}}} 
    	 &log2\_max\_frame\_num\_minus4 & 11 & 13 &Parsing \\ \cline{3-6}
    	 & & pic\_order\_cnt\_type & 2 & 3 &Parsing \\ \cline{3-6}
    	 & &log2\_max\_pic\_order\_cnt\_lsb\_minus4  & 11 & 13 &Parsing \\ \cline{3-6}
    	 & & pic\_width\_in\_mbs\_minus1  & 139 & 256 & Reconstruction \\ \cline{3-6}
    	 & &frame\_cropping\_flag  & 8 & 8 & Reconstruction \\ \cline{3-6}
    	 & &frame\_crop\_right\_offset & 8 & 8 & Reconstruction \\ \cline{2-6} \cline{2-6}
    	 & \parbox[t]{2mm}{\multirow{4}{*}{\rotatebox[origin=c]{90}{PPS}}} 
    	 &entropy\_coding\_mode\_flag& 2 & 2 & Parsing \\ \cline{3-6}
    	 & &transform\_8x8\_mode\_flag & 2 & 2 & Reconstruction\\ \cline{3-6}
    	 & &deblocking\_filter\_control\_present\_flag  & 2 & 2 &Parsing\\ \cline{3-6}
    	 & &pic\_init\_qp\_minus26  & 40 & 52 & Reconstruction\\ \hline 
    	 
    	  \hline
    	  
    	 \parbox[t]{2mm}{\multirow{5}{*}{\rotatebox[origin=c]{90}{Invariant}}} 
    	 & \parbox[t]{2mm}{\multirow{3}{*}{\rotatebox[origin=c]{90}{SPS}}} 
    	 &bit\_depth\_luma\_minus8& 2 & 7 & Parsing \\ \cline{3-6}
    	 & &bit\_depth\_chroma\_minus8  & 2 & 7 & Parsing\\ \cline{3-6}
    	 & &frame\_crop\_left\_offset & 1 & 8 & Reconstruction\\ \cline{2-6}  \cline{2-6}
    	 &\parbox[t]{2mm}{\multirow{2}{*}{\rotatebox[origin=c]{90}{PPS}}} 
    	 &num\_slice\_groups\_minus1  & 1 & 8 & Parsing \\ \cline{3-6}
    	 & &redundant\_pic\_cnt\_present\_flag & 1 & 2 & Parsing\\ \hline 
    	\end{tabular}
	}
\end{table}

\subsection{Invariant Parameters}

The second group includes parameters that are observed to be invariant across the design set.
It overall includes 13 parameters from SPS and 13 from PPS.
It is possible that some of these parameters induce the same decoding behavior as the core parameters.
Our tests on these 26 parameters revealed that a mismatch in the values of five parameters that denote the number of slice groups, presence of redundant pictures, bit depths of luma and chroma samples, and the left cropping amount of a picture also cause decoding failures, as shown in the lower part of Table \ref{tab:parameters}.
However, overall, we encountered only two videos where the depth of the samples of luma and chroma arrays were different (0.04\% of the videos) with all other parameters set to fixed values.

\subsection{Interchangeable Parameters}

These are the group of parameters whose values set at the encoder can be changed without a decoding failure.  
That is, these parameters accept values that supersede actual encoding values while yielding an acceptable decoding output.
Our tests identified 27 parameters that fall in this category. 
Some of these parameters specify the required capabilities at the decoder, such as profiles, levels, and constraint sets.
Those parameters essentially inform the target decoder about the encoding complexity and the needed processing power as well as the bandwidth, resolution and memory requirements. 
In other words, these parameters are not directly used during decoding, and a decoder with sufficient resources can support all settings regardless of the set values.
Similarly, some of the parameters affect the quality of the reconstructed picture such as the height of a picture which may result in stretching or cropping of frames when incorrect. 
Another group of parameters are related to reconstruction of non-IDR frames, such as those related to the reference frames used in construction of the other frames.
Regardless of the values of these parameters, the IDR frame can be decoded successfully. 
However, to be able to recover non-IDR frames they must also be assigned valid values, such as setting the number of reference frames to the highest possible value.

\section{Header Generation }
\label{sec:generation}

We now describe our method for generating an SPS and PPS that can decode a given H.264 coded video file fragment based on our earlier findings.
The crux of our method lies in determining the core parameter values by avoiding a brute force search while setting the invariant parameters to their observed values and the interchangeable parameters to their most flexible and encompassing settings.  
Since this has to be realized through a search, the objective of our method is to determine the values of the missing SPS and PPS parameters in as few trials as possible.
Since each IDR starts with an I frame and the inter-frame coded pictures (P and B frames) cannot be recovered without it, we focus on the ability to recover I frames. However, it must be noted that the values in SPS and PPS do not typically change over a video. Therefore, once an I frame is reconstructed successfully, the same parameters can be used to decode the subsequent frames in an IDR\footnote{It must be noted that to decode the P and B frames \texttt{max\_num\_ref\_frames} parameter must be set to the highest possible value of 16.}.

\begin{figure*}[htbp]
 \centering
  \includegraphics[width=0.9\textwidth]{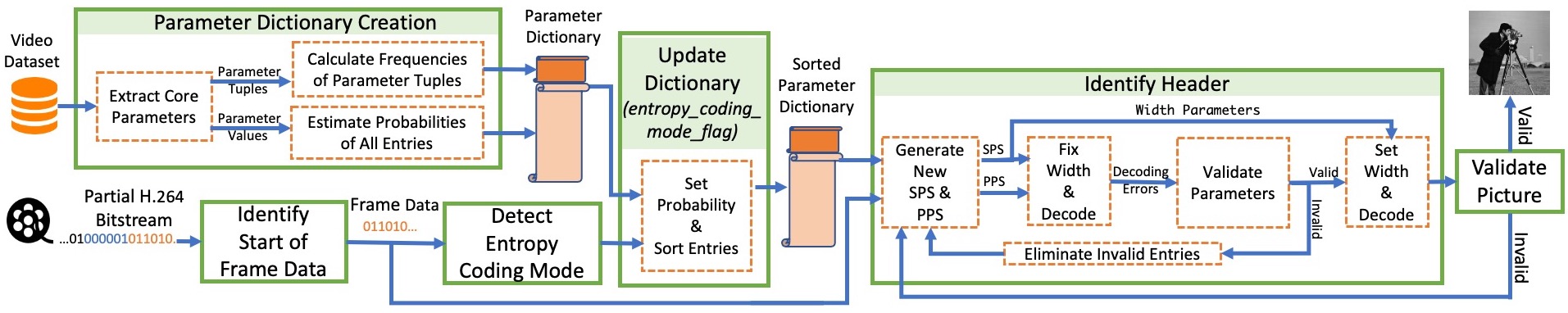}
  \caption{Overview of the header-generation method. SPS and PPS headers are generated based on sorted entries
  obtained from videos in the design set. A search is then performed by iterating over the entries. Through detecting entropy coding mode and utilizing decoder error log messages, invalid headers are eliminated to improve efficiency.}
  \label{fig:pipeline}
\end{figure*}

The flow chart for the overall SPS and PPS sequence header generation algorithm is displayed in Fig. \ref{fig:pipeline}. 
Our method essentially takes as input encoded I frame data and a dictionary of coding parameters and determines the critical parameters needed for decoding. 
The dictionary includes parameter tuples initially sorted based on their observed frequency in the design set, which are further weighted during the search. To avoid an exhaustive search, our method incorporates a learning approach that classifies encoded data and utilizes a mapping between the encountered decoding error messages and parameter settings.
Below we describe the operation of our method in detail.

\subsection{Identifying the Start of Frame Data}
The start of coded frame data can be identified through presence of specific byte patterns included in the beginning of NAL units and the MP4 headers \cite{na2013frame,alghafli2016identification}.
An H.264 coded bitstream is essentially a stream of NAL units separated by a start code prefix, and 
each NAL unit starts with a one-byte unit identifier composed of a zero bit followed by a two-bit NAL reference identification field, and a five-bit NAL unit type.
To identify NAL units in MPEG-4 Visual and H.264 Annex-B formatted files, Na {\em et al.} \cite{na2013frame} proposed searching for start codes \texttt{0x00000001} or \texttt{0x000001} in the bitstream and verifying that they are followed by a header identifier.
In \cite{alghafli2016identification}, Alghafli {\em et al.} further expanded on this approach by considering the MP4 file format in which the four bytes preceding each header identifier are used to store the length of that NAL unit.
To also exploit this, they proposed first detecting all potential header identifiers in a forensic image and identifying actual NAL units by validating that each unit is followed by another header identifier.
To reduce false positive detections, \cite{NFI,lewis2012reconstructing} proposed performing semantic checks to ensure that the structure of NAL units conform to the standard.
In our experiments, we used the first approach to identify IDR headers in coded video files.

\subsection{Parameter Dictionary Creation}
\label{sec:dictionary}

The parameter dictionary is a collection of 63-tuples with each entry $H=(p_1,p_2,\ldots,p_{63})$ representing a realization of all parameters in SPS and PPS. 
Each parameter $p_i\in \mathcal{P}_i$ takes value from a set of possible values in $\mathcal{P}_i$ as defined in the last column of Table \ref{SPS-PPS-Headers} 
in the Appendix. 
Although the encoding process involves a large degree of freedom in the choice of parameters, in practice, what we identify as core parameters determine the complexity of generating SPS and PPS headers blindly.
Therefore, in our search, we assume that only core parameters are unknown while the remaining ones are fixed.
In this regard, the invariant parameters are set in accordance with their encountered values in the design set, and each interchangeable parameter is set to the master value that supersedes other values when decoding.

Overall the dictionary contains around 3.5 billion entries considering possible values for the 10 core parameters. 
Header entries are initially sorted in order of decreasing priority based on two criteria.
The first criterion prioritizes the combination of core parameter values seen in the design set.
Out of the 5,115 unique SPS and PPS headers that cover the design set, we observed 2,663 unique 10-tuples.
Therefore, the first 2.6K header entries incorporate these values sorted based on their frequency in the design set.
The second criterion determines the sorting of subsequent entries based on frequency of each parameter value. 
Since core parameters are all independent from each other, the rank of each header is determined based on its estimated encounter probability computed as multiplication 
of marginal probabilities of each parameter.
That is, considering a generic header $H$ with parameters $p_i=x_i$, for $x_i \in \mathcal{P}_i$, $H$'s rank in the dictionary is determined based on the probability
$\prod_i Pr(p_i=x_i)$ where each probability term represents the normalized occurrence frequency of parameter values in the design set.
Since other than core parameters all parameter values are set to predetermined values, their probabilities are considered as one, and hence they don't 
affect the ranking. 
For those parameter values not seen in the design set, such as several width values, their probabilities are set to a fixed but small value so 
that the resulting header probability is non-zero.

\subsection{Entropy Coding-Mode Detection}
\label{sec:entropymode}

Entropy coding is the last step of encoding.
Therefore, the parameter that identifies the coding method, whether CABAC or CAVLC, is the most important one as the decoder starts interpreting data accordingly. 
In more than 5K unique headers encountered in the design set, we observed that around two thirds used CABAC coding. 
Although newer cameras prefer using CABAC due to its coding efficiency, it is plausible that both methods are commonly used in practice.
Therefore, the ability to infer the type of entropy coding directly from the coded video sequence data will reduce the computational complexity of search by almost one half.

This capability, however, depends on the presence of identifiable differences in the coded bitstream.
The two methods are indeed different in their operations. 
Most notably, CABAC performs entropy coding over all coded elements such as reference frame index, motion vectors, and residual data.
This allows CABAC to perform better modeling of symbol probabilities. 
As a result, one should expect to observe more or less a uniform distribution along the coded sequence.
CAVLC, in contrast, codes only residual data in a context-adaptive manner and other coded elements are coded using Exponential-Golomb codes,
which have a very regular construction with each codeword starting with a prefix of zeros.
Hence, this interleaving of different codes may be expected to introduce deviations from uniformity. 
Our examination of the CAVLC coded data indeed showed that \texttt{0x00} and \texttt{0xFF} values appear more frequently in the byte sequence.

To differentiate between CABAC and CAVLC, we built a simple classifier aimed at exploiting these operational differences.
With this objective, we used Shannon entropy and 11 other features derived from byte frequencies. 
These include two Boolean features testing if the maximum frequency is due to byte-values  \texttt{0x00} and \texttt{0xFF}, the dispersion coefficient computed as the ratio of variance to the mean, the number of byte-values whose frequencies are 1.5 times more than the mean frequency of all byte values as well as six ratios.
The latter six features are computed by dividing \texttt{0x00} byte frequency, \texttt{0xFF} byte frequency, and the maximum frequency by the average and minimum frequencies observed in the entropy coded frame data.

\begin{figure}[!]
  \centering
  \includegraphics[width=0.85\columnwidth]{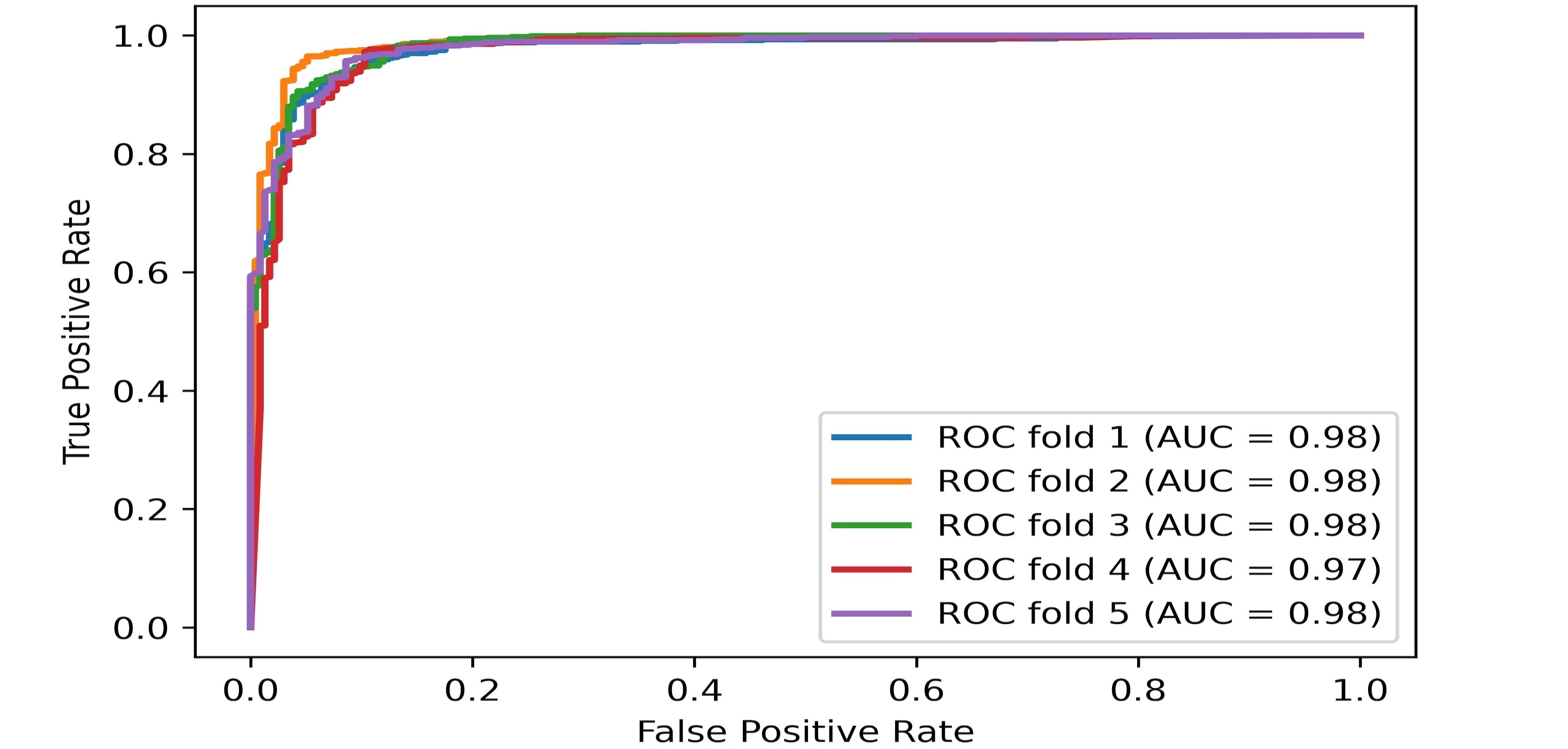}
  \caption{Receiver Operating Curve (ROC) plots generated by five-fold cross-validation to assess the accuracy of entropy coding mode detection using 11 byte frequency features to train a random forest classifier.}
  \label{fig:entropyModel}
\end{figure}

These features are used to build a random forest classifier.
The accuracy of the classifier is determined by performing five-fold cross-validation on a set of I frames encoded by one of the 5,115 SPS and PPS combinations of which 1,197 were coded using CAVLC and the remaining 3,918 using CABAC.
Area Under the Curve (AUC) scores computed for each test fold and the final AUC score, obtained as an average across all folds, are found to be 97-98\%, as shown in Fig. \ref{fig:entropyModel}.
Since the two entropy coding methods can be quite accurately discriminated, the sorting of header entries in the dictionary must take this into account.

\subsection{Updating Dictionary}
\label{sec:width}

The headers in the dictionary are initially sorted based on frequencies of 10-tuples and individual parameter values. 
The above approach essentially builds a binary classifier that predicts the likelihood of the two entropy coding methods, thereby reducing the search space by one parameter. 
Denoting the probability of CABAC coding by $P_e$ and CAVLC coding by $1-P_e$, the dictionary can be updated by re-sorting its entries.
Accordingly, the encounter probability of the first 2.6K entries is multiplied by either $P_e$ or $1-P_e$ depending on the value of the entropy coding-mode parameter.
For the remaining entries the individual probability of encoding type measured across the design set is substituted with classifier's confidence in its prediction ({\em i.e.,} $P_e$ or $1-P_e$), and they are re-sorted within themselves according to their newly computed encounter probability. 
Then, the coded video sequence is decoded in order using headers in the updated dictionary until decoding succeeds.

\subsection{Header Identification}

When decoding a given video sequence data, if the decoder is not correctly initialized decoding eventually fails. 
In the case of the FFMPEG tool, this failure is implicit as the decoder persistently attempts to decode each subsequent picture until it reaches the end of coded data. 
To increase the efficiency of search, we use other supplementary information.
Many widely used applications, such as FFMPEG, have well-designed built-in logs. 
These application logs record important events and provide critical information about the state of the decoder when it fails. 
Our method exploits these error messages both to identify decoding failures and to investigate the mapping between error messages and correct parameter values. 

During decoding, when the assumed parameter values mismatch the actual values used for encoding, several error messages that relate to a missing (top or left) block are logged.
These errors essentially indicate that the decoder is unable to determine the reference block needed during the decoding of the current block.
Our examination of the error patterns indeed revealed that a \textit{top block unavailable error} indicates that at least one of the core parameters is incorrect.
We also identified that \textit{left block unavailable error} indicates an incorrect picture width value.
Essentially, setting an incorrect value for the frame width results in misplacement of decoded picture blocks. 
More specifically, when the selected width value is smaller than the actual value, some picture blocks will inadvertently be carried over to the next row of blocks.
The first misplaced block in that new row will likely raise this error as there will be no blocks to the left of it.

To exploit this behavior when exploring the parameter space, we set the picture width to the smallest possible value of one without cropping.
(It must be noted that core parameters also include a cropping flag bit to indicate if a picture needs to be cropped and the amount of cropping along with the picture width. The former two are set to zeros.)
Since frame resolution for most videos will be higher, this type of error can be utilized as a condition to test the case when all parameters but the width are correctly determined. 
We must note here that the alternative of setting the width to the highest value will result in stacking of many blocks in a single row which is more likely to result in a reference to a non-existing top block.
Therefore, this setting cannot be used as a condition to test correctness of other core parameters.  

Our tests indeed verified that a left block non-present error is always encountered when the width is set to a smaller value. 
However, it is not exclusive to this setting alone and mismatching values in other core parameters also trigger it.
Therefore, when this error is encountered the correctness of the width has to be tested by substituting the value of one with actual width values in the header entries.
It must also be added that with extremely small possibility such an error may not be encountered when none of the misplaced blocks are predicted from their left neighboring blocks during encoding. 
In this unlikely case, decoding will succeed but the picture will be rendered at an incorrect width.

Overall, these two types of errors can be utilized to increase the speed of the search for encoding parameters considerably. 
Essentially, if a top-block cannot be found (when width is set to one), it indicates that at least one of the remaining core parameters is incorrect. 
Therefore, all header entries including those core parameters can be excluded from the search.
Similarly, when a missing left-block error is received, it is likely that all parameters except for the width have been correctly identified.
In that case, rather than trying all possible width values, we prioritize the width values observed in the design set.

\subsection{Picture Validation}

As the last step of our method, we verify whether a reconstructed picture actually exhibits characteristics of real images.
In our tests, we utilized the correlation between the pictures generated using the identified header and the original header used for encoding.
(In a few cases, we noticed that the actual video picture was an empty black frame. 
To take this case into account we additionally checked if the difference between the two pictures are non-zero.)
In practice, however, one needs to decide only based on the decoder output.
This can simply be determined based on the absence of any decoding errors.
Further, our decoding attempts show that when the decoder fails, in most cases a picture cannot be reconstructed, and in the rare cases that a picture is erroneously constructed, it can be easily distinguished from real pictures.

\begin{figure}[htbp]
  \begin{minipage}[b]{0.24\columnwidth}
    \centering
    \includegraphics[width=1\columnwidth]{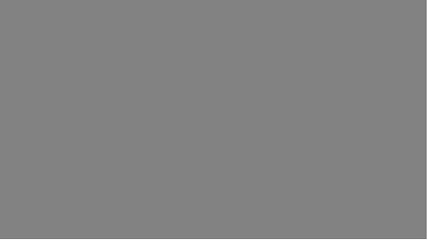}
    \centerline{\scriptsize{(a)}}
  \end{minipage}
  \begin{minipage}[b]{0.24\columnwidth}
    \centering
    \includegraphics[width=1\columnwidth]{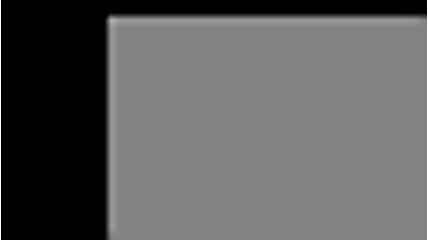}
    \centerline{\scriptsize{(b)}}
  \end{minipage}
  \begin{minipage}[b]{0.24\columnwidth}
    \centering
    \includegraphics[width=1\columnwidth]{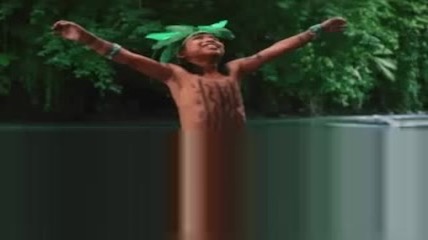}
    \centerline{\scriptsize{(c)}}
  \end{minipage}
    \begin{minipage}[b]{0.24\columnwidth}
    \centering
    \includegraphics[width=1\columnwidth]{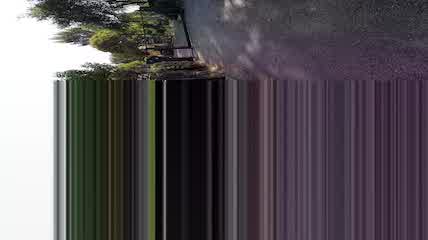}
    \centerline{\scriptsize{(d)}}
  \end{minipage}
  \caption{\textcolor{black}{Examples of pictures generated when parameters were not correct: (a) the gray screen picture that is encountered most frequently, (b) erroneously reconstructed picture, (c) and (d) decoding with incorrect picture height.}}
  \label{fig:erroneous}
\end{figure}

\textcolor{black}{Figures \ref{fig:erroneous}(a) and \ref{fig:erroneous}(b) show two examples where decoding failed but pictures were nevertheless generated. 
In contrast, Figs. \ref{fig:erroneous}(c) and \ref{fig:erroneous}(d) provide examples of decoding where only picture height was incorrect. 
It can be seen that in the latter two pictures, the video context is still discernible.}
During our tests, where we performed tens of millions of decoding attempts, we determined that most erroneously constructed frames yield correlation values less than 0.1, with the highest observed value being 0.25.
Hence for the more general case, this issue can be addressed through building classification models that represent statistical properties of natural images and incorrectly decoded pictures as done in \cite{uzun2015carving}. 

\section{Evaluation}
\label{sec:Evaluation}
In assessing the efficiency of our video sequence header-generation method, we determine the number of trials it takes to identify a valid header entry using our parameter dictionary as described in Sec. \ref{sec:generation}.
We examine how entropy coding-mode detection and the use of error messages to validate core parameters improve the efficiency of search compared to going over dictionary entries one by one in the order in which they appear.
The conventional approach underlying the commercial tool Defraser \cite{NFI} based on stitching previously seen SPS and PPS headers as a whole is also implemented for comparison purposes. 
We will refer to this as the header-stitching method.
For this, we created another dictionary that includes the unique combination of SPS and PPS headers sorted with respect to their occurrence frequency in the design set. 
This dictionary overall includes 7,006 headers due to inclusion of optional VUI parameters in some of the SPS NAL units.
Measurements are performed separately on all videos in both the design and test sets. 

\subsection{Experiments on Design Set}
\label{sec:evaldesign}

This test setting is considered to determine the best achievable performance as we finetune the search method using aggregate statistics of parameter values obtained over the design set.
In our tests, we selected one coded IDR frame data from 5,115 randomly selected videos that are encoded using a unique combination of SPS and PPS units encountered in the design set.
We then attempted to decode each frame data using five methods.
These include the header-generation method; 
the search over core parameters separately incorporated with decoding error messages and entropy coding-mode detection;
the search over core parameters without the additional improvements guiding the search;
and the  header-stitching method.
It must be noted that this measurement setting reflects the best case for the header-stitching method.

\begin{figure}[!]
  \centering
  \includegraphics[width=0.85\columnwidth]{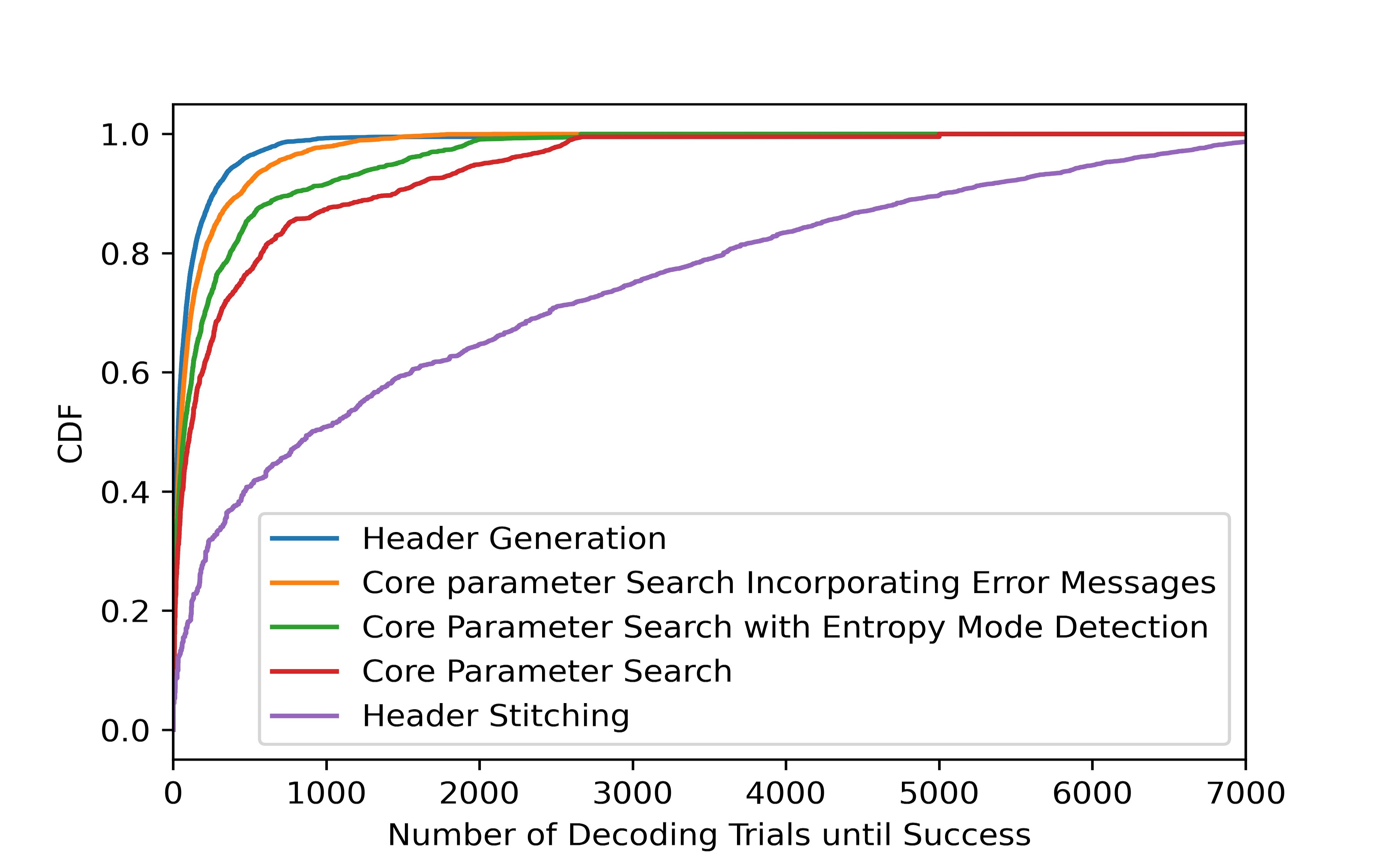}
  \caption{The CDF of the number of trials it takes before successfully decoding coded frame data of 5,115 videos from the design set with a unique SPS and PPS combination.} 
  \label{fig:design}
\end{figure}

Figure \ref{fig:design} shows the cumulative distribution function (CDF) of the number of trials it takes to successfully decode each of the frames taken from the 5,115 videos.
As can be seen in the CDF plots, after 500 trials, the ratio of correctly decoded videos reaches 96.4\% for the header-generation method whereas the other four methods can only decode  91.9\%, 85.9\%, 77.0\%, and 40.8\% of the videos at this level, respectively.
Even at 10 trials, header-generation method is able to decode 26\% of the videos in comparison to 22\%, 21\%, 14\%, and 5\% of other four methods.

The difference between the header-generation and header-scotching methods can be mainly attributed to two main factors. 
First, with header-stitching, headers are sorted based on the encounter probability of SPS and PPS headers. 
In contrast, by only considering core parameters, the header-generation approach can more reliably sort the headers. 
Second factor stems from the fact that our parameter dictionary excludes all optional parameters related to post-decoding processing stages.
This effectively reduces the search space for the header-generation method to 2.6K entries as compared to 7K entries for the header-stitching method.

Overall, our search results show that it takes on average 116, 147, 263, 405, and 1,792 trials to decode each video using the above five methods in their respective order.
These results show that the header-generation method offers an improvement by a factor of 15 in the search speed over the header-stitching method.
Similarly, the use of decoding error messages and the entropy coding-mode detection combined together provides
close four times improvement in the search speed over core parameter search alone.
They also demonstrate that incorporation of the entropy coding-mode classifier and the error messages to identify encoding parameters improves the search speed by almost 4 times when compared to core parameter search. 
Identifying the entropy coding mode effectively reduces the space by one parameter. 
More importantly, since picture width varies over a wide range of values, by isolating it from the search ({\em i.e.,} setting it to one) the search becomes even more efficient.

\subsection{Experiments on Test Set}

To validate and compare the performance of header-generation methods on previously unseen videos, we use the test set.
It must be noted that earlier tests were performed on a subset of videos that include a sample of each unique SPS and PPS header combination observed in the design set.
Our analyses on 
VISION dataset show that different models tend to use different SPS and PPS headers and within each model
SPS and PPS headers remain largely the same.
Over 31 different models from 11 brands in the VISION dataset, only one model possesses two pairs of SPS and PPS headers, while all other models possess one pair of headers.
Overall, these 31 models possess 21 unique pairs of SPS and PPS headers, showing that parameter sets are largely distinctive and invariant.
Therefore, the higher diversity of PPS and SPS headers potentially indicates the use of a large number of source cameras and video editing tools in the design set.
By picking one video from each unique header combination, those results essentially demonstrate the generalization capability of methods across different models and editing tools.
To obtain a more realistic setting, the test set includes randomly selected videos to reflect the real-world prevalence of encoding settings.
That is, we impose no restriction on the number of videos that may have the same SPS and PPS headers.

The videos in the test set are further divided into two subsets depending on which \texttt{lbry.tv} user account they are obtained from.
The first subset includes 37,460 videos uploaded by 3,979 users that overlap with the 11,034 users spanning the design set.
Assuming these videos will exhibit similar provenance characteristics, our goal is to evaluate the degree of variability in SPS and PPS headers. 
The second subset includes 18,561 videos associated with 5,840 non-overlapping user accounts and will highlight the generalization capability of our method.
The two subsets will be referred to as overlapping and non-overlapping test videos. 
Although the non-overlapping test set contains videos captured by fewer users when compared to the design set, it nevertheless contains a large number of previously unseen SPS and PPS headers.
Considering that videos in the VISION dataset yielded 21 SPS and 11 PPS headers in total, the non-overlapping test set can be considered quite diverse.
Similar to the above test setting, we extracted one coded IDR frame data from each video for decoding. 

An important difference between using design and test sets in the experiments is that our header-generation method is guaranteed (except for two videos) to find a working header among the first 2,663 dictionary entries.
In this test setting, however, since new SPS and PPS headers are likely to be encountered the search may well expand into lower ranked entries of the dictionary that are sorted based on combined prevalence of individual parameter values.
Essentially, the thoroughness of our dictionary will be determined based on whether it contains all required header entries and the average rank of those header entries in the dictionary.

Our examination of videos in the test set revealed that they include 4,154 SPS and 301 PPS headers out of which 3,025 SPS and 61 PPS headers do not appear in the design set.
More relevantly, these headers include 1,483 unique 10-tuples of core parameters, 818 of which are also among the 2,663 tuples seen in the design set. 
Further, these 818 instances cover 55,094 videos in the test set, indicating that their headers will be identified in less than 2,663 trials, leaving only 927 videos for a lengthier search.
However, it must be remembered that quantization parameters need not be exact and many of the 927 videos could still be decoded with some of the available parameters.

Of more immediate concern is whether any of the parameters that are deemed to be invariant take alternative values as those SPS and PPS headers cannot be composited using our parameter dictionary.
Our examination of the unique SPS and PPS headers in the test set revealed that bit depths of luma and chroma samples in 3 videos are different. 
(It must be noted that a similar variation in the two parameters was also observed in two videos of the design set.)
This essentially shows that our parameter groupings are quite general and that our header-generation method can composite almost all headers encountered among the test videos.

\begin{figure}[!]
    \centering
  \includegraphics[width=0.85\columnwidth]{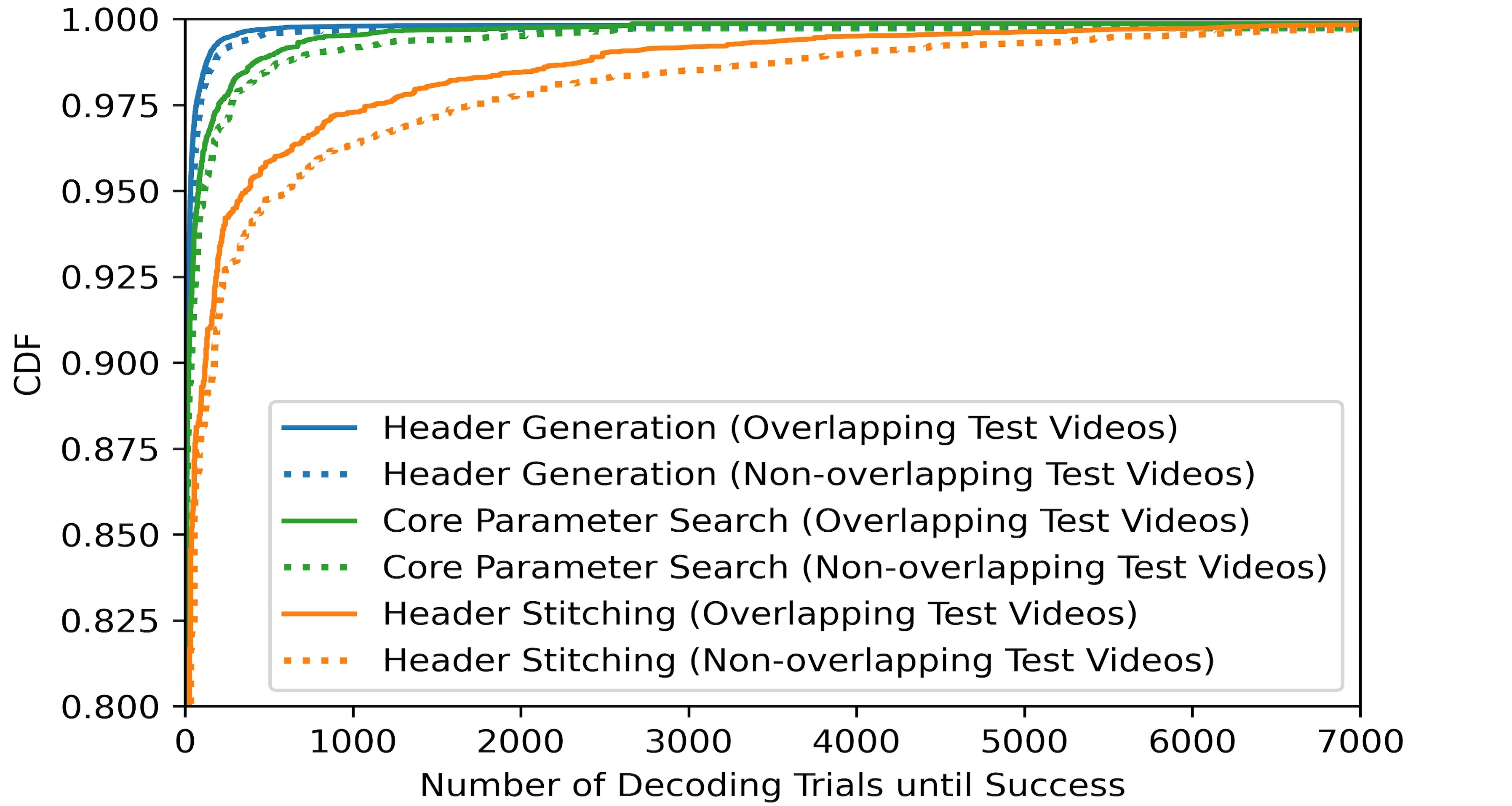}
  \caption{CDF of the number of trials it takes to successfully decode a video frame data in the overlapping test set.}
  \label{fig:test1}
\end{figure}

Figure \ref{fig:test1} displays the CDF for the number of trials it takes to identify a header to decode overlapping and non-overlapping test videos. 
Accordingly, after 10 and 500 decoding trials the header-generation method can decode 87.8\% and 99.7\% of the videos in the overlapping video test set.
Considering the core parameter search method, the numbers, respectively, reduce to 84.6\% and 98.9\%.
The header-stitching method performs substantially worse than the other two methods, decoding only 59.7\% of videos after 10 trials and 95.5\% of videos after 500 trials. 
Fig. \ref{fig:test1} also provides the decoding performance of the three methods on the non-overlapping video test set (dotted curves).
As expected the performance deteriorated for all methods due to inclusion of videos with unseen headers. 
For the header-generation method the performance gap is marginal at best as the number of decodable videos after 10 and 500 trials, respectively, include 86.6\% and 99.6\% of videos.
This gap, however, increasingly widens for the other methods. 
The difference between the CDFs obtained on overlapping and non-overlapping test videos is essentially due to the prevalence of parameter values in the two subsets, 
which ultimately affects the order of header entries in the dictionary.
Overall, these results indicate the robustness of the header-generation approach in handling more diverse encoding settings.

Search results also show that 58 and 81 videos in overlapping and non-overlapping test sets are at the tails of the distributions.
That is, parameters comprising these header entries are ranked beyond the first 2.6K entries in the dictionary.
Not including these videos, it takes on average $10.9$ and $12.4$ trials, respectively, for the header-generation method to identify correct headers 
on the overlapping and non-overlapping videos.
Corresponding averages are determined to be $22.7$ and $29.8$ trials for the core parameter search method in contrast to 
$108.4$ and $151.9$ trials for the header-stitching method.
Overall, a very large majority of videos in the test set can be decoded in $11.3$ trials using our header-generation method.
This number increases to $24.9$ trials when the search is performed over core parameters and $122.6$ trials when header-stitching is performed.

It must be noted that the average number of decoding attempts per video on the test set, {\em i.e.,} Fig. \ref{fig:test1}, is much smaller than found earlier on the design set, {\em i.e.,} Fig. \ref{fig:design}.
This is because the 5,115 videos used for experiments in Sec. \ref{sec:evaldesign} are all encoded using different headers.
Essentially, by assuming each header is equally likely to be encountered, this setting provides the worst-case header generation complexity in terms of the number of decoding trials for the design set.
By contrast, videos in the test set reflect the real-world prevalence of video headers.
That is, some video headers are more frequently encountered than others. 
Since those frequently occurring headers are likely to be higher ranked entries in the parameter dictionary, they can be decoded in fewer trials, thereby reducing the overall average.
The results on the test set also show that our parameter dictionary is able to capture general encoding characteristics and verifies our intuition that encoding parameters do not vary significantly across videos.

We further analyzed the 58 and 81 videos in the two subsets that require large numbers of decoding trials.
A search over the coding parameters in the header dictionary will, respectively, take on average 1.75M and 1.09M attempts to identify the correct headers 
for those videos.
To determine the improvement provided by the header-generation method, we continued the search for 50 of the 139 videos that are ranked higher in the parameter dictionary based on estimated encounter probabilities.
It took the header-generation method 2,459 attempts on average to identify these headers compared to 9,778 attempts required by the core parameter search.
The improvement by a factor close to four shows that the header-generation method becomes more effective in identification of less prevalent encoding settings due to its ability to eliminate invalid headers.
It must be noted that the header-stitching method cannot decode any of these videos.
We also examined how unique these videos are in terms of their encoding settings by determining the total number of videos encoded using the same 
parameter values.
Our examination revealed that each encoding setting is used in coding of 1.76 videos on average with only 9 headers appearing in more than one video.  
This shows that the ranks of these headers in the dictionary well reflect their prevalence in practice.

\subsection{Complexity}

The theoretical complexity of all methods, as evaluated in Figs. \ref{fig:design} and \ref{fig:test1}, are linear in the number parameter settings (or headers) available in the dictionary for decoding.
It must be noted that our method sorts parameters settings in relation to their real-world prevalence observed in the design set, thereby making the actual decoding time probabilistic depending on the camera used for capturing a video.
Since both our header-generation method and the header-stitching method rely on the use of a header dictionary, the time complexities of both methods are the same from this perspective.
Our tests performed on a workstation with an Intel Xeon (2.4 GHz) processor and 16 GB memory running the Ubuntu operating system show that for a given coded IDR frame data, validating a pair of SPS and PPS headers takes on average 0.11 seconds.

Our header-generation method, however, incurs additional complexity due to the entropy coding-mode detection and the incorporation of the decoding error messages to the analysis. 
In this regard, the entropy coding-mode detection step is performed only once for a given coded frame data and can be performed very fast. 
Therefore, its impact is insignificant.
The error message analysis involves string comparison and access to the header dictionary to eliminate invalid entries which requires lightweight computation compared to the other operations such as decoder initialization.
Our measurements show that the overall time overhead of the header-generation method in decoding a frame in comparison to the header-stitching method is less than 5\% for validating a pair of SPS and PPS headers.
It must be noted that parallelization of search steps will increase the efficiency of these methods linearly with the degree of parallelism.

\section{Discussion}
\label{sec:Discussion}

Video coding offers a large freedom in the choice of parameters.  
Our analysis of the encoding parameters of more than a hundred thousand videos, however, show that most encoding settings cluster tightly in the parameter space.
In this regard, we determined that a large number of encoding parameters are either invariant or not very critical to reconstructing pictures. 
This is an indication that certain encoding settings are commonly preferred by devices and video processing tools. 
Results show that our header-generation method requires 54.6\% fewer trials to identify the encoding parameters in comparison to a search over core parameters using our parameter dictionary. 
The improvement becomes much more apparent when compared with the conventional header-stitching approach, which uses the previously seen SPS and PPS headers as a whole, 
where the number of trials is reduced by 90.8\%.

More distinctively, our method relies on error messages logged by the decoder when using incorrect parameters.
This is an important side information about the state of the decoder.
Therefore, our first intuition was to create a mapping between parameter values and 67 log messages observed. 
However, our exploratory analysis showed that mismatches in any of the core parameters yield a similar pattern of errors due to the similar impact they induce on the operation of the decoder.
Our future studies will explore more advanced learning approaches to distinguish potential differences in messages.

Generalization of our carving capability to arbitrary videos is another important concern.
In this regard, the size of the design set and the diversity of SPS and PPS headers contained therein determine the real-world performance of the header-generation method as these factors influence the sorting of entries in the parameter dictionary. 
The test set used for evaluation of our method included a total of 56K videos with 3,025 SPS and 61 PPS headers that did not appear in the design set,  including 48K videos.
We determined that the search complexity is relatively high only for 139 videos in the test set, essentially requiring more than a million trials to generate their headers.  
This indicates that the design set is sufficiently comprehensive and the reported performance results to hold in practice.

In practice, a very large number of videos are generated and shared through social media platforms, such as YouTube, Instagram, Tiktok, {\em etc}. 
These videos are typically re-encoded to meet different viewing options and bandwidth requirements in a platform dependent manner.
Therefore, encoding settings should exhibit little variation across platform videos.
In fact, our exploratory analysis on videos obtained from YouTube and Instagram at varying resolutions shows that their encoding settings are covered by our parameter dictionary among the top 600 entries. 
For TikTok videos, we determined that some of them have width values not seen in our design set.
These findings show that our header-generation method, with some further finetuning of our parameter dictionary, will potentially be effective on a large set of videos distributed through the Internet.

\section{\textcolor{black}{Conclusion and Future Work}}
\label{sec:Conclusion}
\textcolor{black}{
In this work, we address the problem of automatically identifying the encoding parameters needed to decode a video file fragment.
This is an important and missing capability in video file carving, where SPS and PPS headers containing these parameters comprise only a very small part of a video file and may be missing for several reasons. 
Essentially, our approach composites the missing SPS and PPS NAL units using a parameter dictionary that reflects the real-world prevalence of each parameter.
Rather than performing an exhaustive search over all parameters, our method categorizes parameters into three groups and focuses on only those that are absolutely necessary and highly variant. 
Our method also evaluates statistical characteristics of coded bitstream to determine the entropy coding mode and, more importantly, makes use of error messages encountered during decoding to identify correct parameter values, thereby further reducing the search complexity.}

\textcolor{black}{
A key objective of this line of research is the generalization of this capability to other encoding formats while reducing the level of domain expertise needed to generate
such decoders. 
This requires devising approaches that can learn the core parameters for those codecs in a blind manner, 
and determining valid, decoding parameter values associated with a file fragment in a more dynamic manner rather than using a precomputed parameter dictionary.
Our future work will focus on building this capability by incorporating more advanced learning approaches to the proposed parameter search method.
}

\appendix
\label{sec:appendix}

The parameters that comprise SPS and PPS sequence headers are given in Table \ref{SPS-PPS-Headers}. 
The first column of the table shows the header type as SPS or PPS, and the second column provides our designation of each parameter in terms of its criticality for decoding.
The third and fourth columns provide the name and the defined range of values for each parameter as specified by the standard.
We refer the reader to \cite{tsbmail} for a detailed description of each parameter.
The possible range of values for each parameter is given in the last column.
The grouping of parameters is observed to be consistent over both the design and test sets.
Only in the context of invariant parameters, five videos (two in the design set and three in the test set) out of more than 100K videos are determined to take alternative values in representing bit depths of luma and chroma samples.
Due to low probability of encountering this setting, we considered both parameters to be invariant in value.

It must also be noted that SPS has an optional part that includes the video usability information (VUI) parameters which provide additional information about higher-level
properties of video content such as aspect ratio, color space, chroma location, bitstream restrictions, timing etc. 
Since VUI parameters are not involved in decoding of individual pictures, our header generation method disregards VUI parameters.

\vspace{-0.1cm}
\begin{table}[!h]
	\centering
	\caption{Categorization of SPS and PPS Parameters and Defined Range of Values }
	\label{SPS-PPS-Headers}
	
	\resizebox{\linewidth}{!}{
	\begin{tabular}{|c|c|c|c|}
		\hline
	Header&Group & Variable Name  & Possible Value \\ \hline
		\hline

\parbox[t]{2mm}{\multirow{30}{*}{\rotatebox[origin=c]{90}{SPS}}} & \parbox[t]{2mm}{\multirow{6}{*}{\rotatebox[origin=c]{90}{Core}}} & pic\_width\_in\_mbs\_minus1  &  0-255   \\ \cline{3-4}

& & frame\_cropping\_flag   & 0-7   \\\cline{3-4}

& & frame\_cropping\_right   &  0,1    \\\cline{3-4}  

& &log2\_max\_frame\_num\_minus4  &  0-12   \\\cline{3-4}

& & pic\_order \_cnt\_type  &  0-2  \\\cline{3-4}

& & log2\_max\_pic\_order\_cnt\_lsb\_minus4   &  0-12      \\\cline{2-4}

& \parbox[t]{2mm}{\multirow{12}{*}{\rotatebox[origin=c]{90}{Invariant}}} &    chroma\_format\_idc  & 0-3  \\  \cline{3-4}

& &separate\_colour\_plane\_flag  &  0,1 \\\cline{3-4}

& &\multirow{1}{*}{\parbox{3cm}{\centering bit\_depth\_luma\_minus8 }}  &  \multirow{1}{*}{\parbox{1cm}{\centering 0-6}}  \\ \cline{3-4}

& &\multirow{1}{*}{\parbox{3cm}{\centering bit\_depth\_chroma\_minus8 }}  &  \multirow{1}{*}{\parbox{1cm}{\centering 0-6}} \\ \cline{3-4}

& &frame\_cropping\_top-left  &  0-7   \\\cline{3-4}

& &qpprime\_y\_zero\_transform\_bypass\_flag  & 0,1 \\\cline{3-4}

& &delta\_pic\_order\_always\_zero\_flag   &  0,1   \\\cline{3-4}

& & \multirow{2}{*}{\parbox{3cm}{\centering offset\_for\_(non)\_ref\_pic or top\_to\_bottom\_field }}  &  \multirow{2}{*}{\parbox{2cm}{\centering (-2$^{31}$+1)-(2$^{31}$-1)}} \\ & & &  \\ \cline{3-4}

& & num\_ref\_frames\_in\_pic\_order\_cnt\_cycle   &   0-255 \\\cline{3-4}

& &gaps\_in\_frame\_num\_value\_allowed\_flag   & 0,1  \\\cline{3-4}

& &mb\_adaptive\_frame\_field\_flag  & 0,1  \\\cline{2-4}

&   \parbox[t]{2mm}{\multirow{12}{*}{\rotatebox[origin=c]{90}{Interchangeable}}} &  seq\_parameter\_ set\_id   &  0-31 \\  \cline{3-4}

& 		&profile\_idc & 18 profiles \\ \cline{3-4}
& 		&level\_idc & 0-255 \\ \cline{3-4}
&    & constraint\_set (0,1,2,3,4,5)\_flag  &   0-1 for each    \\ \cline{3-4}

& & seq\_scaling\_matrix\_present\_flag   &   0-16  \\\cline{3-4}

& & pic\_height\_in\_mbs\_minus1  &  0-255  \\ \cline{3-4}

& &frame\_mbs\_only\_flag   &  0,1   \\ \cline{3-4}

& &frame\_cropping\_bottom  &  0-7    \\\cline{3-4}

& &seq\_scaling\_list\_present\_flag[ i ]   &  0,1  \\\cline{3-4}

& &max\_num\_ref\_frames  &  0-16 \\\cline{3-4}

& &direct\_8x8\_inference\_flag  & 0,1\\\cline{3-4}

& &vui\_parameters\_present\_flag   &  0,1 \\\hline
\hline
		
\parbox[t]{2mm}{\multirow{25}{*}{\rotatebox[origin=c]{90}{PPS}}} & \parbox[t]{2mm}{\multirow{4}{*}{\rotatebox[origin=c]{90}{Core}}} & \multirow{1}{*}{\parbox{3cm}{\centering entropy\_coding\_mode\_flag  }}  &  \multirow{1}{*}{\parbox{1cm}{\centering 0,1}} \\  \cline{3-4}

& & \multirow{1}{*}{\parbox{3cm}{\centering pic\_init\_qp\_minus26 }}  &  \multirow{1}{*}{\parbox{1cm}{\centering (-25)-26}}    \\  \cline{3-4}

& & \multirow{1}{*}{\parbox{3cm}{\centering transform\_8x8\_mode\_flag}}  &  \multirow{1}{*}{\parbox{1cm}{\centering 0,1}}     \\ \cline{3-4}

& & deblocking\_filter\_control\_present\_flag   &  0,1  \\ \cline{2-4}

& \parbox[t]{2mm}{\multirow{12}{*}{\rotatebox[origin=c]{90}{Invariant}}} &bottom\_field\_pic\_order \_in\_frame\_present\_flag   &  0,1  \\\cline{3-4}

& & \multirow{1}{*}{\parbox{3cm}{\centering num\_slice\_groups\_minus1  }}  &  \multirow{1}{*}{\parbox{1cm}{\centering 0-7}} \\ \cline{3-4}

& &\multirow{1}{*}{\parbox{3cm}{\centering slice\_group\_map\_type  }}  &  \multirow{1}{*}{\parbox{1cm}{\centering 0-6}}  \\ \cline{3-4}

& &\multirow{1}{*}{\parbox{3cm}{\centering run\_length\_minus1  }}  &  \multirow{1}{*}{\parbox{1cm}{\centering 0-256}}  \\ \cline{3-4}

& &\multirow{1}{*}{\parbox{3cm}{\centering top\_left[i] bottom\_right[i] }}  &  \multirow{1}{*}{\parbox{1cm}{\centering 0,1}}  \\ \cline{3-4}

& &slice\_group\_change\_direction\_flag   &   0,1   \\ \cline{3-4}

& &slice\_group\_change\_rate\_minus1  &   0-$2^{16}$   \\ \cline{3-4}

& &pic\_size\_in\_map\_units\_minus1   & 0-$2^{16}$ \\ \cline{3-4}

& &\multirow{1}{*}{\parbox{3cm}{\centering slice\_group\_id[i] }}  &  \multirow{1}{*}{\parbox{1cm}{\centering 0,1}}  \\ \cline{3-4}

& &\multirow{1}{*}{\parbox{3cm}{\centering chroma\_qp\_index\_offset }}  &  \multirow{1}{*}{\parbox{1cm}{\centering (-12)-12}}  \\ \cline{3-4}

& &\multirow{1}{*}{\parbox{3cm}{\centering constrained\_intra\_pred\_flag}}  &  \multirow{1}{*}{\parbox{1cm}{\centering 0,1}}  \\ \cline{3-4}

& &redundant\_pic\_cnt\_present\_flag  &   0,1  \\ \cline{2-4}

& \parbox[t]{2mm}{\multirow{9}{*}{\rotatebox[origin=c]{90}{Interchangeable}}} & \multirow{1}{*}{\parbox{3cm}{\centering pic\_parameter\_set\_id }}  &  \multirow{1}{*}{\parbox{1cm}{\centering 0-255}} \\ \cline{3-4}

& &\multirow{1}{*}{\parbox{3cm}{\centering seq\_parameter\_set\_id }}  &  \multirow{1}{*}{\parbox{1cm}{\centering 0-31}}  \\ \cline{3-4}

& &num\_ref\_idx\_l(0-1)\_ default\_active\_minus1   &   0-31 \\ \cline{3-4}

& &\multirow{1}{*}{\parbox{3cm}{\centering weighted\_pred\_flag }}  &  \multirow{1}{*}{\parbox{1cm}{\centering 0,1}}   \\ \cline{3-4}

& &\multirow{1}{*}{\parbox{3cm}{\centering weighted\_bipred\_idc }}  &  \multirow{1}{*}{\parbox{1cm}{\centering 0-2}}   \\ \cline{3-4}

& &\multirow{1}{*}{\parbox{3cm}{\centering pic\_init\_qs\_minus26 }}  &  \multirow{1}{*}{\parbox{1cm}{\centering (-25)-26}}  \\  \cline{3-4}

& &pic\_scaling\_matrix\_present\_flag   &   0-16   \\\cline{3-4}

& &pic\_scaling\_list\_present\_flag[ i ]   & 0-256  \\\cline{3-4}

& &second\_chroma\_qp\_index\_offset   &  (-12)-12   \\\hline

	\end{tabular}
	}
\end{table}

\bibliographystyle{IEEEtran}
\bibliography{bibfile}

\begin{thebibliography}{10}
\providecommand{\url}[1]{#1}
\csname url@samestyle\endcsname
\providecommand{\newblock}{\relax}
\providecommand{\bibinfo}[2]{#2}
\providecommand{\BIBentrySTDinterwordspacing}{\spaceskip=0pt\relax}
\providecommand{\BIBentryALTinterwordstretchfactor}{4}
\providecommand{\BIBentryALTinterwordspacing}{\spaceskip=\fontdimen2\font plus
\BIBentryALTinterwordstretchfactor\fontdimen3\font minus
  \fontdimen4\font\relax}
\providecommand{\BIBforeignlanguage}[2]{{%
\expandafter\ifx\csname l@#1\endcsname\relax
\typeout{** WARNING: IEEEtran.bst: No hyphenation pattern has been}%
\typeout{** loaded for the language `#1'. Using the pattern for}%
\typeout{** the default language instead.}%
\else
\language=\csname l@#1\endcsname
\fi
#2}}
\providecommand{\BIBdecl}{\relax}
\BIBdecl

\bibitem{garfinkel2007carving}
S.~L. Garfinkel, ``Carving contiguous and fragmented files with fast object
  validation,'' \emph{Digital Investigation}, vol.~4, pp. 2--12, 2007.

\bibitem{poisel2011advanced}
R.~Poisel, S.~Tjoa, and P.~Tavolato, ``Advanced file carving approaches for
  multimedia files.'' \emph{J. Wirel. Mob. Networks Ubiquitous Comput.
  Dependable Appl.}, vol.~2, no.~4, pp. 42--58, 2011.

\bibitem{alghafli2019}
K.~Alghafli, C.~Y. Yeun, and E.~Damiani, ``Techniques for measuring the
  probability of adjacency between carved video fragments: The vidcarve
  approach,'' \emph{IEEE Transactions on Sustainable Computing}, 2019.

\bibitem{twoStage}
J.~Fang, G.~Xi, R.~Li, Q.~Chen, P.~Lin, S.~Li, Z.~L. Jiang, and S.-M. Yiu,
  ``Coarse-to-fine two-stage semantic video carving approach in digital
  forensics,'' \emph{Computers \& Security}, vol.~97, p. 101942, 2020.

\bibitem{lewis2012reconstructing}
A.~B. Lewis, ``Reconstructing compressed photo and video data,'' University of
  Cambridge, Computer Laboratory, Tech. Rep., 2012.

\bibitem{park2014data}
J.~Park and S.~Lee, ``Data fragment forensics for embedded {DVR} systems,''
  \emph{Digital Investigation}, vol.~11, no.~3, pp. 187--200, 2014.

\bibitem{NFI}
E.~Casey and R.~Zoun, ``Design tradeoffs for developing fragmented video
  carving tools,'' \emph{Digital Investigation}, vol.~11, pp. S30--S39, 2014.

\bibitem{na2013frame}
G.-H. Na, K.-S. Shim, K.-W. Moon, S.~G. Kong, E.-S. Kim, and J.~Lee,
  ``Frame-based recovery of corrupted video files using video codec
  specifications,'' \emph{IEEE Transactions on Image Processing}, vol.~23,
  no.~2, pp. 517--526, 2013.

\bibitem{pereira2002mpeg}
F.~Pereira and T.~Ebrahimi, \emph{The MPEG-4 book}.\hskip 1em plus 0.5em minus
  0.4em\relax Prentice Hall Professional, 2002.

\bibitem{tsbmail}
T.~ITU, ``Advanced video coding for generic audiovisual services,'' \emph{ITU-T
  Recommendation {H.264}}, 2003.

\bibitem{yannikos2013automating}
Y.~Yannikos, N.~Ashraf, M.~Steinebach, and C.~Winter, ``Automating video file
  carving and content identification,'' in \emph{IFIP International Conference
  on Digital Forensics}.\hskip 1em plus 0.5em minus 0.4em\relax Springer, 2013,
  pp. 195--212.

\bibitem{ei2019}
E.~Tandogan, E.~Altinisik, S.~Sarimurat, and H.~T. Sencar, ``Tackling in-camera
  downsizing for reliable camera {ID} verification.''\hskip 1em plus 0.5em
  minus 0.4em\relax Society for Imaging Science and Technology, 2019.

\bibitem{reconstructHeader}
K.~Sheng, X.~Liao, Q.~Zhang, J.~Qu \emph{et~al.}, ``Video forensic of
  fragmented video based on {H.264/AVC} video compression standard,'' in
  \emph{2014 International Conference on Mechatronics, Electronic, Industrial
  and Control Engineering (MEIC-14)}.\hskip 1em plus 0.5em minus 0.4em\relax
  Atlantis Press, 2014.

\bibitem{uzun2015carving}
E.~Uzun and H.~T. Sencar, ``Carving orphaned {JPEG} file fragments,''
  \emph{IEEE Transactions on Information Forensics and Security}, vol.~10,
  no.~8, pp. 1549--1563, 2015.

\bibitem{uzun2019jpg}
------, ``{JpgScraper}: An advanced carver for jpeg files,'' \emph{IEEE
  Transactions on Information Forensics and Security}, vol.~15, pp. 1846--1857,
  2019.

\bibitem{gloe2014forensic}
T.~Gloe, A.~Fischer, and M.~Kirchner, ``Forensic analysis of video file
  formats,'' \emph{Digital Investigation}, vol.~11, pp. S68--S76, 2014.

\bibitem{iuliani2018video}
M.~Iuliani, D.~Shullani, M.~Fontani, S.~Meucci, and A.~Piva, ``A video forensic
  framework for the unsupervised analysis of mp4-like file container,''
  \emph{IEEE Transactions on Information Forensics and Security}, vol.~14,
  no.~3, pp. 635--645, 2018.

\bibitem{lopez2020digital}
R.~R. L{\'o}pez, E.~A. Luengo, A.~L.~S. Orozco, and L.~J.~G. Villalba,
  ``Digital video source identification based on container’s structure
  analysis,'' \emph{IEEE Access}, vol.~8, pp. 36\,363--36\,375, 2020.

\bibitem{huaman2020authentication}
C.~Q. Huam{\'a}n, A.~L.~S. Orozco, and L.~J.~G. Villalba, ``Authentication and
  integrity of smartphone videos through multimedia container structure
  analysis,'' \emph{Future Generation Computer Systems}, vol. 108, pp. 15--33,
  2020.

\bibitem{song2016integrity}
J.~Song, K.~Lee, W.~Y. Lee, and H.~Lee, ``Integrity verification of the ordered
  data structures in manipulated video content,'' \emph{Digital Investigation},
  vol.~18, pp. 1--7, 2016.

\bibitem{guera2019we}
D.~G{\"u}era, S.~Baireddy, P.~Bestagini, S.~Tubaro, and E.~J. Delp, ``We need
  no pixels: Video manipulation detection using stream descriptors,''
  \emph{arXiv preprint arXiv:1906.08743}, 2019.

\bibitem{mpeg1}
D.~Le~Gall, ``Mpeg: A video compression standard for multimedia applications,''
  \emph{Communications of the ACM}, vol.~34, no.~4, pp. 46--58, 1991.

\bibitem{H263}
G.~Cote, B.~Erol, M.~Gallant, and F.~Kossentini, ``H. 263+: Video coding at low
  bit rates,'' \emph{IEEE Transactions on circuits and systems for video
  technology}, vol.~8, no.~7, pp. 849--866, 1998.

\bibitem{H264}
T.~Wiegand, G.~J. Sullivan, G.~Bjontegaard, and A.~Luthra, ``Overview of the
  {H.264/AVC} video coding standard,'' \emph{IEEE Transactions on circuits and
  systems for video technology}, vol.~13, no.~7, pp. 560--576, 2003.

\bibitem{H265}
G.~J. {Sullivan}, J.~{Ohm}, W.~{Han}, and T.~{Wiegand}, ``Overview of the high
  efficiency video coding (hevc) standard,'' \emph{IEEE Transactions on
  Circuits and Systems for Video Technology}, vol.~22, no.~12, pp. 1649--1668,
  2012.

\bibitem{bitmovin}
``Video developer survey 2019,''
  \url{https://go.bitmovin.com/video-developer-survey-2019}, (Accessed on
  08/29/2021).

\bibitem{lbry}
\BIBentryALTinterwordspacing
``lbry.tv.'' [Online]. Available: \url{https://lbry.tv/}
\BIBentrySTDinterwordspacing

\bibitem{socrates}
C.~Galdi, F.~Hartung, and J.-L. Dugelay, ``Socrates: A database of realistic
  data for source camera recognition on smartphones.'' in \emph{ICPRAM}, 2019,
  pp. 648--655.

\bibitem{vision}
D.~Shullani, M.~Fontani, M.~Iuliani, O.~Al~Shaya, and A.~Piva, ``Vision: a
  video and image dataset for source identification,'' \emph{EURASIP Journal on
  Information Security}, vol. 2017, no.~1, pp. 1--16, 2017.

\bibitem{h264bitstream}
``Github - aizvorski/h264bitstream,''
  \url{https://github.com/aizvorski/h264bitstream}, (Accessed on 04/07/2021).

\bibitem{alghafli2016identification}
K.~Alghafli and T.~Martin, ``Identification and recovery of video fragments for
  forensics file carving,'' in \emph{2016 11th International Conference for
  Internet Technology and Secured Transactions (ICITST)}.\hskip 1em plus 0.5em
  minus 0.4em\relax IEEE, 2016, pp. 267--272.

\end{thebibliography}

\end{document}